\documentclass[english,aps,preprint,showpacs]{revtex4}
\usepackage[T1]{fontenc}
\usepackage[latin9]{inputenc}
\usepackage{esint}
\usepackage{graphicx}

\makeatletter
\@ifundefined{textcolor}{}
{%
 \definecolor{BLACK}{gray}{0}
 \definecolor{WHITE}{gray}{1}
 \definecolor{RED}{rgb}{1,0,0}
 \definecolor{GREEN}{rgb}{0,1,0}
 \definecolor{BLUE}{rgb}{0,0,1}
 \definecolor{CYAN}{cmyk}{1,0,0,0}
 \definecolor{MAGENTA}{cmyk}{0,1,0,0}
 \definecolor{YELLOW}{cmyk}{0,0,1,0}
 }

\makeatother

\usepackage{babel}

\begin{document}

\title{$\mathcal{PT}$ Symmetry and the Sign Problem}

\author{Peter N. Meisinger, Michael C. Ogilvie and Timothy D. Wiser}

\affiliation{ Dept. of Physics
Washington University
St. Louis, MO 63130 USA}

\begin{abstract}
Generalized $\mathcal{PT}$ symmetry provides crucial insight
into the sign problem for two classes of models. 
In the case of quantum statistical models
at non-zero chemical potential,
the free energy density is directly related to the ground
state energy of a non-Hermitian, but 
generalized $\mathcal{PT}$-symmetric Hamiltonian.
There is a corresponding class of $\mathcal{PT}$-symmetric
classical statistical mechanics models with non-Hermitian
transfer matrices.
For both quantum and classical models,
the class of models with generalized $\mathcal{PT}$ symmetry is precisely
the class where the complex weight problem can be reduced
to real weights, \emph{i.e.}, a sign problem. The spatial two-point
functions of such models can exhibit three different behaviors: exponential
decay, oscillatory decay, and periodic behavior. The latter two regions
are associated with $\mathcal{PT}$ symmetry breaking, where a Hamiltonian
or transfer matrix has complex conjugate pairs of eigenvalues. The transition
to a spatially modulated phase is associated with $\mathcal{PT}$
symmetry breaking of the ground state, and is generically a first-order
transition. In the region where $\mathcal{PT}$ symmetry is unbroken,
the sign problem can always be solved in principle. Moreover, there
are models with $\mathcal{PT}$ symmetry which can be simulated for
all parameter values, including cases where $\mathcal{PT}$
symmetry is broken.
\end{abstract}


% 11.30.Er	Charge conjugation, parity, time reversal, and other discrete symmetries

% 11.10.Wx	Finite-temperature field theory
 
% 64.60.Bd	General theory of phase transitions
 
% 64.60.De	Statistical mechanics of model systems (Ising model, Potts model, field-theory models, Monte Carlo techniques, etc.)

\maketitle

\section{The complex weight problem}

The sign problem occurs in many branches of theoretical
physics, including particle physics, many-body theory, statistical
physics and condensed matter theory. The problem arises when the expected
value of an observable is computed as a sum over non-positive
weights, and perhaps should be called more generally the complex weight
problem. For example, in lattice QCD, the introduction of a non-zero
chemical potential makes the quark contribution to the functional
determinant complex \cite{deForcrand:2010ys,Lombardo:2008sc}.
Because the powerful simulation techniques of
lattice gauge theory rely on a positive weight within the functional
integral, progress in simulating QCD at finite density has been meager
relative to what has been achieved in other aspects of QCD. Similar
problems occur in other areas of physics, such as the Hubbard model
away from half-filling \cite{Loh:1990zz} and systems
with topological charges \cite{Cox:1999nt}.

We will show that $\mathcal{PT}$ symmetry is a powerful tool for analyzing
the sign problem. In particular, $\mathcal{PT}$-symmetric systems are precisely
the class of systems where the complex weight problem can be reduced
to a sign problem, i.e., positive and negative weights. Furthermore,
$\mathcal{PT}$ symmetry provides a natural classification scheme for the
oscillatory behaviors observed in correlation functions
in liquid and crystalline phases, giving us a simple,
unified picture of behavior across many areas of physics. 
Finally,
$\mathcal{PT}$ symmetry gives us important information about the solvability of
the sign problem, and a framework for future progress in work on particular
systems.

The fundamental importance of $\mathcal{PT}$ symmetry was first pointed out by
Bender and Boettcher in their seminal work on quantum-mechanical 
models \cite{Bender:1998ke}.
Their work grew out of the observation that the Hamiltonian
\begin{equation}H=p^{2}+igx^{3}.\end{equation}
which arises naturally in the study of the Lee-Yang theory of phase
transtions \cite{Lee:1952ig,Fisher:1978pf,Cardy:1985yy,Cardy:1989fw,Zamolodchikov:1989cf,Yurov:1989yu}, 
has only real eigenvalues. Bender and Boettcher observed
that the Hamiltonian $H$, while not Hermitian, is invariant under the
simultaneous application of the symmetry operations
parity $\mathcal{P}:\, x\rightarrow-x$ and
time reversal $\mathcal{T}:\, i\rightarrow-i$. 
This symmetry ensures that all eigenvalues of $H$ are either real
or part of a complex pair. The argument is simple: if $H\left|\psi\right\rangle =E\left|\psi\right\rangle $
then $H\mathcal{PT\,\left|\psi\right\rangle =}\mathcal{PT}\, H\left|\psi\right\rangle =\mathcal{PT}\, E\left|\psi\right\rangle =E^{*}\mathcal{PT}\,\left|\psi\right\rangle $. Thus if $E$ is an eigenvalue, $E^*$ is an eigenvalue as well.

\section{Quantum Many-body theory and $\mathcal{PT}$ Symmetry}

We will now show that all quantum many-body problems involving a non-zero
chemical potential may be described in terms of a non-Hermitian Hamiltonian
with generalized $\mathcal{PT}$ symmetry. We start from a theory with a
Hermitian Hamiltonian
$H$ and a conserved global quantum number $N$, obtained from a conserved
current $j_{\mu}$, that commutes with $H$. We assume that $H$ is
Hermitian and invariant under the combined action of time reversal
$\mathcal{T}$ and a charge conjugation $\mathcal{C}$ that reverses
the sign of $j^{\mu}$. We take the number of spatial dimensions to
be $d-1$, and the spatial volume to be $L^{d-1}$. The grand canonical
partition function at temperature $T=\beta^{-1}$ and chemical potential
$\mu$ is given by $Z=Tr\left[\exp\left(-\beta H+\beta\mu N\right)\right]$.
If $Z$ is written as
a Euclidean path integral, the time component of the current $j^{0}$
will Wick rotate to $ij^{d}$, while the chemical potential $\mu$
does not change. This leads directly to a non-positive weight in the
path integral, and is the origin of the sign problem in
finite density calculations: probabilistic methods do not
work. The Euclidean space Lagrangian density may be written as $\mathcal{L}-i\mu j^{d}$
where $\mathcal{L}$ is the Euclidean Lagrangian for $\mu=0$; $\mathcal{L}-i\mu j^{d}$
is complex. The nature of the problem is changed by changing the direction
of Euclidean time, so that we are now considering a problem at zero
temperature with one compact spatial dimension of circumference $\beta$.
Upon returning to Minkowski space, $j^{d}$ does not rotate. We pick,
say, the $1$ direction to be the new time direction and take the
new inverse temperature $L$ to satisfy $L\gg\beta$. When $\mu=0$,
the original Hamiltonian is unchanged, but the partition function
is now given by\begin{equation}
Z=Tr\left[e^{-LH_{\beta}}\right]\end{equation}
where\begin{equation}
H_{\beta}=H-i\mu\int d^{d-1}x\, j^{_{d}}.\end{equation}
The new Hamiltonian $H_{\beta}$ is non-Hermitian, but possesses a
generalized $\mathcal{PT}$ symmetry, where the role of $\mathcal{P}$
is played by the charge conjugation operator $\mathcal{C}$ that changes
the sign of $j^{0}$ and $N$. Under the combined action of $\mathcal{CT}$,
$j^{d}\rightarrow-j^{d}$ and $i\rightarrow-i$, leaving the Hamiltonian
$H_{PT}$ invariant. 
If we introduce the operator $H_{L}=H-\mu N$,
we have the relation\begin{equation}
Z= Tr\left[e^{-\beta H_{L}}\right]=Tr\left[e^{-LH_{\beta}}\right]\end{equation}
under the space-time transformation that exchanges directions
$1$ and $d$. Note that $Z$ is obtained from $H_{L}$ by a sum over
all eigenstates, but is dominated by the  ground state of $H_{\beta}$
in the limit of large $L$.

\begin{figure}
\includegraphics[width=4.5in]{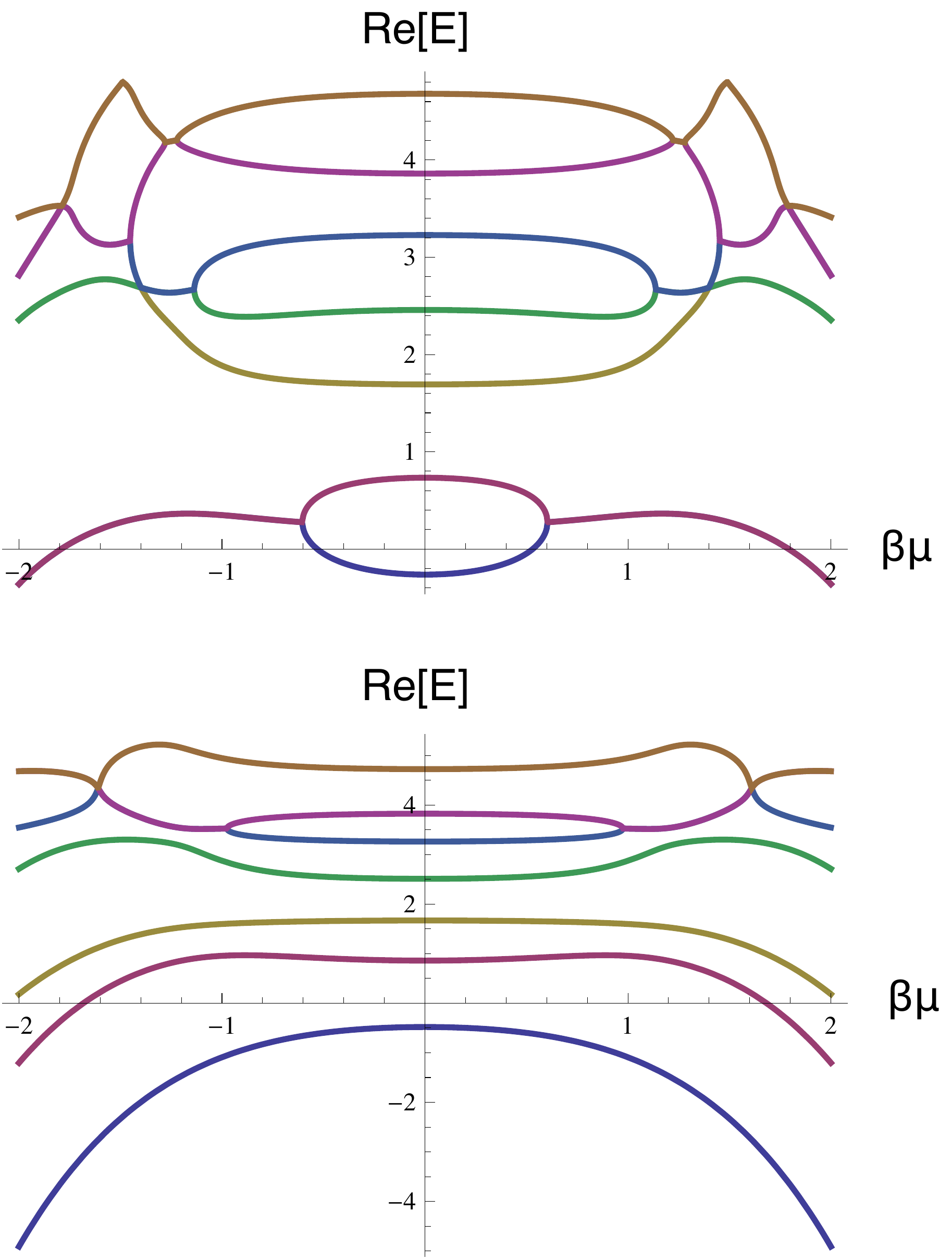}
\caption{The real part of the $SU(3)$ Hamiltonian $H_\beta$ 
as a function of $\beta\mu$. The upper graph is for periodic boundary conditions for the heavy quarks, while the lower graph is for antiperiodic periodic boundary conditions. 
The energy has been scaled such that $g^2\beta/2$ is set equal to $1$.
}\label{fig:su3}
\end{figure}

We have previously given an explicit example of the construction
of $H_{\beta}$ for the case of QCD with static quarks at finite density
\cite{Ogilvie:2008zt}.
After the static quarks are integrated out of the functional integral,
the effective action has the form\begin{equation}
S_{eff}=\int d^{d}x\left[\frac{1}{4g^{2}}\left(F_{\mu\nu}^{a}\right)^{2}-h_{F}\left(e^{\beta\mu}Tr_{F}(P)+e^{-\beta\mu}Tr_{F}(P^{+})\right)\right]\end{equation}
where $P$ is the Polyakov loop $P\left(\vec{x}\right)=\mathcal{P}\exp\left[i\int_{0}^{\beta}dtA_{4}\left(\vec{x},t\right)\right]$
and $h_{F}$ is a known function of the temperature and the quark
mass $M$, going to zero as $M$ goes to infinity. In $1+1$ dimensions,
the solution of this model reduces to solving a $\mathcal{PT}$-symmetric
quantum mechanics problem on the gauge group. The Hamiltonian $H_{\beta}$,
obtained from $S_{eff}$, is \begin{equation}
H_{\beta}=\frac{g^{2}\beta}{2}C_{2}-h_{F}\beta\left[e^{\beta\mu}Tr_{F}(P)+e^{-\beta\mu}Tr_{F}(P^{+})\right]\end{equation}
and acts on the gauge-invariant physical states which are class functions
of $P$ obeying $\Psi\left[P\right]=\Psi\left[gPg^{+}\right]$. The
operator $C_{2}$ is the quadratic Casimir operator for the gauge
group, the Laplace-Beltrami operator on the group manifold. The group
characters form an orthonormal basis on the physical Hilbert space:
any gauge-invariant state $\Psi$ can be expanded as
$\Psi\left[P\right]=\sum_{R}a_{R}Tr_{R}\left(P\right)$. The Hamiltonian
$H$ is not Hermitian when $\mu\ne0$, a direct manifestation of the
sign problem, but it is $\mathcal{PT}$ symmetric when $\mathcal{P}$
is taken to be the charge conjugation
operator $\mathcal{C}$ that changes $A_{\mu}$ to $-A_{\mu}$.
In figure \ref{fig:su3}, we show the real part of the eigenvalues of $H_{\beta}$,
measured in units where $g^2\beta/2$  is set to $1$. The overall strength
of the potential term is set by the dimensionless parameter $2h_F/g^2$.
In the upper graph, $2h_F/g^2=0.5$, corresponding to antiperiodic
boundary conditions for the heavy quarks. Note the formation of a
complex conjugate pair of excited state eigenvalues as $\beta\mu$
increases.
It is also of interest to consider the case of periodic boundary conditions
for the heavy quarks, corresponding to $h_F<0$ 
\cite{Myers:2007vc,Unsal:2008ch,Myers:2009df}.
The lower graph shows the energy eigenvalues for $2h_F/g^2=-0.5$,
where the ground state becomes part of a conjugate pair as
$\beta\mu$ increases.
The coalescence of real energy eigenvalues into conjugate pairs
with degenerate real parts 
is typical
of $\mathcal{PT}$-symmetric systems, and is usually referred to as
broken $\mathcal{PT}$ symmetry. Note that this usage is somewhat
different from what is meant by broken symmetry in Hermitian models,
where only the symmetry of the ground state is considered.
Note that for $N\ge3$, the heavy quark finite density problem of
$SU(N)$ gauge theory is in the universality class of the Lee-Yang
problem for $Z(N)$ spin systems \cite{DeGrand:1983fk}.

\section{Models from Statistical Mechanics}

It is therefore not surprising that there is a class of $\mathcal{PT}$-symmetric
$Z(N)$ spin models which are closely related. On each lattice site
$j$ there is a spin $w_{j}$, an element of the group $Z(N)$ which
may be parametrized as $w_{j}=\exp\left(2\pi in_{j}/N\right)$ with
$n_{j}\in\left\{ 0,1,...,N-1\right\} $ defined modulo $N$ so that
$0$ and $N$ are identified. We take the operator $\mathcal{P}$
to be charge conjugation, acting as $n_{j}\rightarrow-n_{j}$, or
equivalently $w_{j}\rightarrow w_{j}^{*}$. The operator $\mathcal{T}$
is again complex conjugation. Although $\mathcal{P}$ and $\mathcal{T}$
have the same effect on the $w_{j}$'s, one is a linear operator and
the other antilinear. We will show below that $\mathcal{P}$ is implemented
as a unitary matrix in the transfer matrix formalism. The Hamiltonian
$H$ is defined by\begin{equation}
-\beta H=\sum_{\left\langle jk\right\rangle }\frac{J}{2}\left(w_{j}w_{k}^{*}+w_{j}^{*}w_{k}\right)+\sum_{j}
\left[ h_R \left( w_{j}+w_{j}^{*}\right) +h_I \left( w_{j}-w_{j}^{*}\right)\right]
\end{equation}
 where $\beta=1/T$, $J$, $h_{I}$ and $h_{I}$ are real and the sum over $\left\langle jk\right\rangle $
represents a sum over nearest-neighbor pairs.
$H$ is trivially
$\mathcal{PT}$-symmetric. This class of models has complex Boltzmann
weights for $N\ge3$ when $h_{I} \ne 0$.

\begin{figure}
\includegraphics[width=5in]{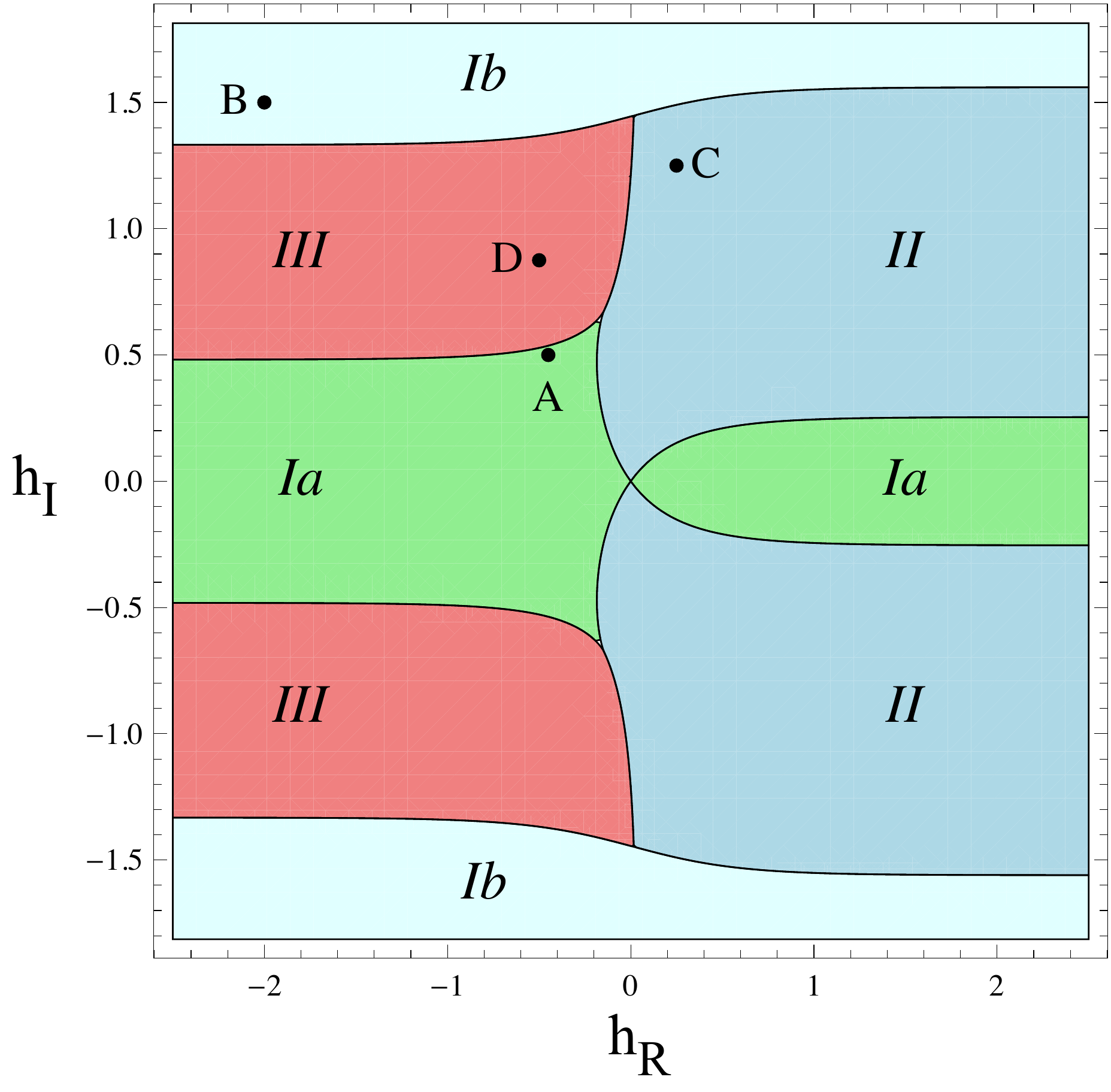}
\caption{Phase diagram for the $d=1$ ${\mathcal PT}$-symmetric 
$Z(3)$ spin model in the $h_R - h_I$ plane at $J = 0.2$.
The interpretation of regions Ia, Ib, II and III are given in the text.}
\label{fig:z3phases}
\end{figure}

\begin{figure}
\includegraphics[width=5in]{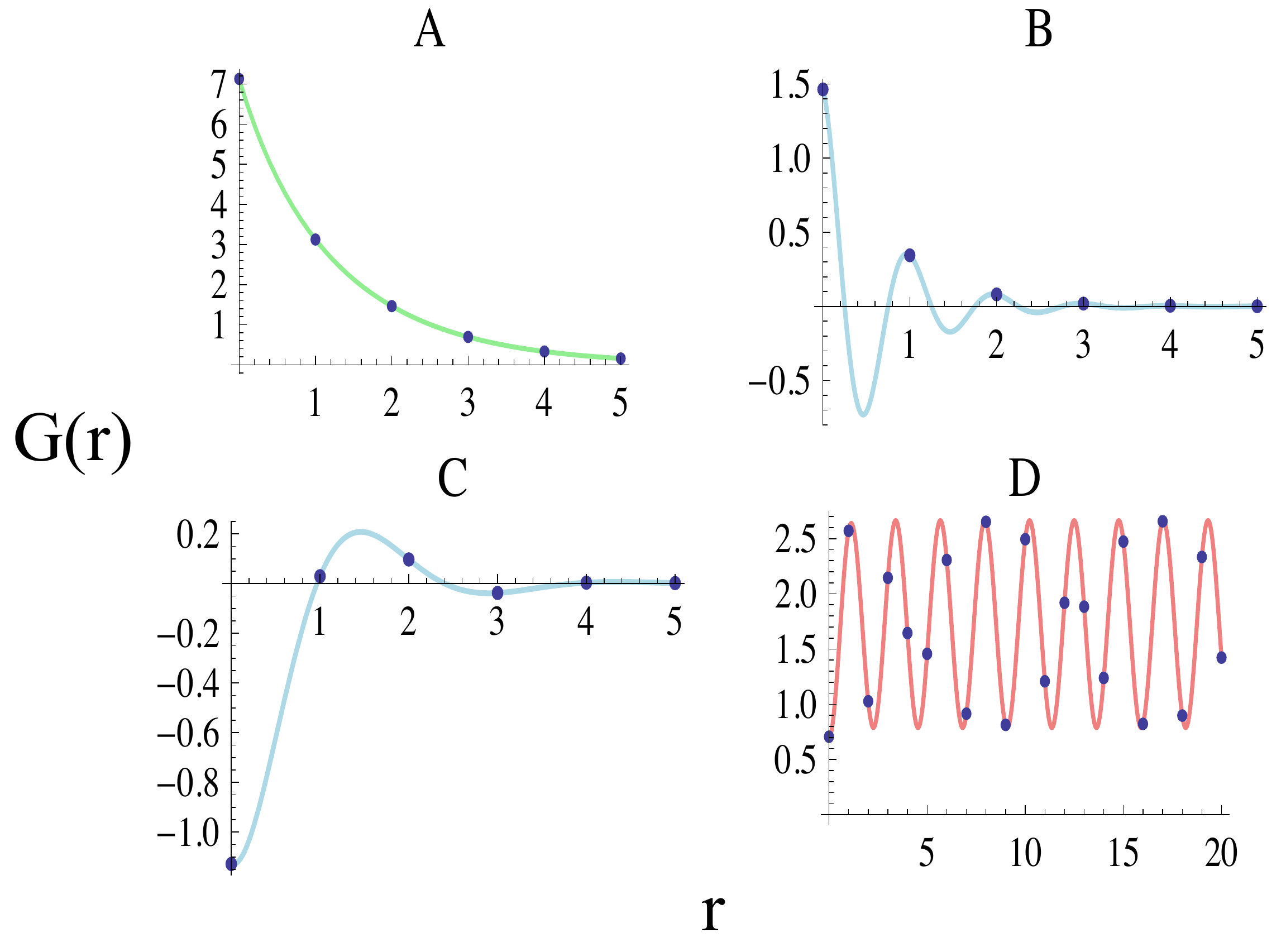}
\caption{The two-point function as a function of lattice spacing for the parameters 
corresponding to points
A, B, C and D in figure \ref{fig:z3phases}.}
\label{fig:z3correlation}
\end{figure}

We illustrate the rich behavior possible in these models using the
case of a $Z(3)$ model in $d=1$. If $h_{I}=0$ , then
the transfer matrix $T$
is Hermitian. When $h_{I}\neq0$, $-\beta H$ is no longer
real and $T$ is no longer Hermitian, but is $\mathcal{PT}$ symmetric.
Figure \ref{fig:z3phases} shows the phase diagram in the $h_{R}-h_{I}$ plane for
$J=0.2$. There are four distinct regions. In region Ia, all three
eigenvalues of the transfer matrix are real and positive. This region
includes the line $h_{I}=0$, and has properties similar to those
found in the Hermitian case. In region Ib, all of the eigenvalues
are real, but at least one of them is negative. In region II, the
eigenvalue of $T$ largest in magnitude is real, but the two other
eigenvalues form a complex conjugate pair. In region III, the two
eigenvalues largest in magnitude form a complex conjugate pair, and
the third, smaller, eigenvalue is real. In both region II and region
III, $\mathcal{PT}$symmetry is broken, but in different ways. Borrowing
the terminology from $\mathcal{PT}$-symmetric quantum mechanics, 
we will describe the behavior
in region III as $\mathcal{PT}$-symmetry breaking of the ground state,
while region II is $\mathcal{PT}$ - symmetry breaking of an excited
state. The behavior of the two-point function
$G\left( \left| j-k \right| \right) =\left< w \left( j \right)w^\dagger \left( k \right) \right>$
 differs substantially
in the three regions. In region I, the two-point function falls off
exponentially. We show typical behavior in region Ia in figure
\ref{fig:z3correlation}
for point A where $\left(h_{R},h_{I}\right)=\left(-0.45,0.5\right)$. 
Similar behavior occurs in region Ib,
as shown in the figure for point B where $\left(h_{R},h_{I}\right)=\left(-2.0,1.5\right)$.
Although the figure shows that the continuation of the two-point function
away from integer values can be negative, note that the values at
integer points are all non-negative. The two-point
function at point C in region II
where $\left(h_{R},h_{I}\right)=\left(0.25,1.25\right)$ 
shows the damped oscillatory behavior associated with $\mathcal{PT}$
breaking in excited states. 
For the point D in region III, where $\left(h_{R},h_{I}\right)=\left(-0.5,0.875\right)$,
the $\mathcal{PT}$ breaking of the ground state leads
to oscillatory behavior of the two-point function in the limit of
large distance.
Note that region III only occurs when $h_{R}$ is negative.
For $h_{R}<0$ and $h_{I}=0$, the spin configurations with lowest
energy have a two-fold degeneracy. With $h_{I}=0$, the ground state
of the transfer matrix
is unique. For the case $h_R<0$, $h_I=0$, and $J$ large,
the splitting of the two lowest eigenvalues of the transfer matrix
in $d=1$ is small.
For sufficiently strong
$h_{I}$, the real parts of the two lowest eigenvalues of $T$ merge,
and $\mathcal{PT}$ symmetry breaking of the ground state occurs.

$\mathcal{PT}$ symmetry is not always manifest in the Hamiltonian or transfer
matrix.
A simple criterion for
a $\mathcal{PT}$-symmetric Hamiltonian $H$ (or transfer matrix $T$)
has been given by Bender and Mannheim \cite{Bender:2009mq}. 
If the characteristic polynomial
$\det\left[H-\lambda I\right]$ has real coefficients, then $H$ has
a generalized $\mathcal{PT}$ symmetry. In the case where $H$ is
real, interesting models arise when $H$ is not symmetric. A notable
example of this behavior is the ANNNI model \cite{Selke:1988zz}. 
The behavior seen above
in the $d=1$ $Z(3)$ model is quite general. There are
generically three regions in $\mathcal{PT}$-symmetric models: 
region I, in which $\mathcal{PT}$ symmetry is unbroken; region II,
in which $\mathcal{PT}$ symmetry is broken by one or more pairs
of excited states becoming complex; and region III, in which $\mathcal{PT}$
symmetry is broken by the ground state becoming complex. In region
II, thermodynamic properties are unaffected, but oscillatory behaviors
appears in correlation functions. In region III, the system is in
a spatially modulated phase. We wish to emphasize that the behavior
of correlation functions seen in regions II and III cannot be obtained
from classical spin models for which $H_{\beta}$ is Hermitian: such
behavior is incompatible with the spectral representation of the the
correlation function for Hermitian theories.

\section{Zeros of the partition function in region 3}

The change from region I to region II is generally not a phase transition
in the thermodynamic limit.
Within a transfer matrix framework, the largest
eigenvalue is unique and real in both regions, so thermodynamic behavior
is smooth. The change only appears in correlation functions. In condensed
matter physics, the locus of points in parameter space where the change
from region I to region II occurs has been known in statistical physics
for some time 
\cite{Fisher:1969yy}
and is often called the disorder line . 
The
transition from region I or II to region III is generically a first-order
phase transition.
There is a general theory of partition function zeros that can be
applied to $\mathcal{PT}$-symmetric lattice models \cite{Biskup:2000xx}. Under
some technical conditions, the partition function in a periodic volume
$V=L^{d}$ can be written as\begin{equation}
Z=\sum_{m}e^{-\beta Vf_{m}}+\mathcal{O}\left(e^{-L/L_{0}}e^{-\beta Vf}\right)\end{equation}
where $f=\min_{m}Re\left[f_{m}\right]$ and $L_{0}$ is of the order
of the largest correlation length of the system. The $f_{m}$'s have
the interpretation of complex free energy densities, and are independent
of $L$. These phases are stable if $Re\left(f_{m}\right)=f$ or metastable
otherwise. The zeros of the partition function are within $\mathcal{O}\left(e^{-L/L_{0}}\right)$
of the solutions of the equations. \begin{eqnarray*}
Re\left(f_{m}\right) & = & Re\left(f_{n}\right)=f\\
Im\left(f_{m}\right) & = & Im\left(f_{n}\right)+\left(2p+1\right)\frac{\pi}{\beta V}\end{eqnarray*}
for some $m\ne n$ and $p\in Z$ . We can apply this directly to region
III, using the representation\begin{equation}
Z=\sum_{r}e^{-LE_{r}}+\sum_{p}\left(e^{-LE_{p}}+e^{-LE_{p}^{*}}\right)\end{equation}
of the partition function. We identify $LE_{0}$ and $LE_{0}^{*}$
as $\beta L^{d}f_{0}$ and $\beta L^{d}f_{0}^{*}$, so that the partition
function has a zero for values of the parameters such that\begin{equation}
\beta\, Im\left[f_{0}\right]=\frac{\left(2p+1\right)\pi}{2V}\end{equation}
 This tells us that the zeros of the partition function lie on the
boundary of region III, defined by $Im\left[f_{0}\right]=0$, in the
limit $V\rightarrow\infty$. As the volume of the system is taken
to infinity, the zeros of the partition function lie asymptotically
on the boundary between phases. Note that this analysis depends on
$L_{0}$ remaining finite. At a 2nd-order transition, $L_{0}$ goes
to infinity and the approximation is invalid.

\section{Towards a solution of the sign problem}

The difficulty presented by the sign problem depends
directly on 
$\mathcal{PT}$ symmetry breaking or its absence.
Mostafazadeh has proven that when $\mathcal{PT}$ symmetry is unbroken
(region I), there is a similarity transformation $S$ that transforms
a $\mathcal{PT}$-symmetric Hamiltonian $H$ into an isospectral Hermitian
Hamiltonian $H_{h}$ via $H_{h}=SHS^{-1}$ \cite{Mostafazadeh:2003gz}. 
This eliminates the sign problem
for $\mathcal{PT}$-symmetric quantum Hamiltonians
throughout region I, if $S$ can be found.
This theorem also applies
to $\mathcal{PT}$-symmetric transfer matrices, but
a further restriction to positive eigenvalues
is necessary
for the elimination of the sign problem.

In regions II and III, the sign problem has an underlying
physical basis. The negative weight
contributions to the partition function $Z$ arise
from the contributions of complex conjugate
eigenvalue pairs associated with $\mathcal{PT}$ symmetry
breaking. It is that breaking that in turn
gives rise to
the oscillatory and damped oscillatory
behavior of two-point functions
characteristic of many physical systems.
However, there are $\mathcal{PT}$ symmetric
models like the ANNNI model \cite{Selke:1988zz} where
the classical Hamiltonian, corresponding to the action in the path
integral formalism, is real.
Such a model can be simulated with no difficulties
of principle throughout its parameter space.
The key point is that the 
the transfer matrix is real but not
symmetric. 
The existence of an antiunitary involution
commuting with the Hamiltonian
implies that there is a basis in which $H$ is real;
 see, {\it e.g.}, \cite{Haake}.
This theorem easily extends to the case of 
those $\mathcal{PT}$-symmetric systems 
for which $\left( \mathcal{PT} \right)^2=1$,
and can be applied to transfer matrices
as well as Hamiltonians.
This suggests the existence of a broad class
of $\mathcal{PT}$-symmetric models which can
be simulated in all three regions,
but the extent of this class is as yet unknown.

%\section{Conclusions}

%In summary, models with generalized $\mathcal{PT}$symmetry are exactly
%the class of models for which the complex weight problem reduces to
%the sign problem. This class includes all quantum statistical models
%with non-zero chemical potential and a large class of classical models
%as well. $\mathcal{PT}$ symmetry breaking is associated with the
%disorder line (in moving from region I to region II) and the onset
%of spatial modulation (region III). Furthermore, $\mathcal{PT}$ symmetry
%provides a physical and mathematical basis for further investigation
%of the sign problem.

\end{document}